\documentclass[10pt]{iopart}
\usepackage{iopams}
\usepackage{cite}
\usepackage{graphicx}
\usepackage{xcolor}
\begin{document}

\title[Controlling electron flow in anisotropic Dirac materials heterojunctions]{Controlling electron flow in anisotropic Dirac materials heterojunctions: A super-diverging lens}

\author{Y. Betancur-Ocampo}
\address{Instituto de Ciencias F\'isicas, Universidad Nacional Aut\'onoma de M\'exico, Cuernavaca, M\'exico}
\address{Departamento de F\'isca Aplicada, Centro de Investigaci\'on y de Estudios Avanzados del IPN, Apartado Postal 73 Cordemex 97310 M\'erida, Yucat\'an, M\'exico}
\ead{ybetancur@icf.unam.mx}
\date{\today}

\begin{abstract}
Ballistic heterojunctions of Dirac materials offer the opportunity of exploring optics-like phenomena in electronic systems. In this paper, a new perfect lens through special positive refraction is predicted with omnidirectional Klein tunneling of massless Dirac fermions. The novel optics component called a super-diverging lens (SDL) is the counterpart of a Veselago lens (VL). The use of SDL and VL creates a device that simulates the ocular vision. This atypical refraction is due to electrons obeying different Snell's laws of pseudo-spin and group velocity in heterojunctions with elliptical Dirac cones. These findings pave the way for an electron elliptical Dirac optics and open up new possibilities for the guiding of electrons.
\end{abstract}

\pacs{73.23.Ad, 73.40.Gk, 73.63.-b}
\submitto{\JPCM}
\vspace{2pc}
\noindent{\it Keywords}: Topological materials, Negative refraction, Klein tunneling, Veselago lens, Heterojunctions, Dirac fermions, Ballistic transport

\maketitle


\section{Introduction} \label{intro}

In the past, the resemblance of photons and electrons has been well used for technological applications where electron microscope is perhaps the most famous example. Nowadays, both entities are closer by the emergence of relativistic materials \cite{XLQi,Kou,Novo,Castro,Wehling,Moore,Armitage}. This new concept in condensed matter has allowed to classify a wide variety of systems, whose excitations present a pseudo-relativistic behavior \cite{XLQi,Kou,Novo,Castro,Wehling,Moore,Armitage}. Thus, graphene $p$-$n$ junction served as platform for the implementation of Klein tunneling \cite{Katsnelson,Kim,Beenakker,Fuchs,Wilmart}, negative refraction of Dirac fermions \cite{HoLee,Chen}, and gate-controlled guiding of electrons \cite{Williams,Rickhaus,Liu}. Electronic components operating as authentic light-geometrical optics systems such as collimators \cite{Pereira2,Park,Louie,Richter}, filters \cite{Jiang,Masum}, Dirac fermion microscopes \cite{Boggild,Hills}, fiber-optic guidings \cite{Williams,Liu}, interferometers \cite{Khan,Rickhaus}, reflectors \cite{Gunlycke} and valley beam splitters \cite{Pomar,Zhai,Stegmann,Charlier,Li} have been proposed. Interesting optical-like phenomena, such as Goos-H\"anchen effect and chiral-dependent Imbert-Fedorov shift, have been studied in Weyl semimetals \cite{Jiang2,Yang}. Currently, an important interest is to use electron optics for controlling valley degree of freedom as conveyor of quantum information \cite{Glattli}. The rapid and simultaneous advances of electron optics and properties of relativistic Dirac materials likely will lead to the realization of concrete technological applications in a near future.

Negative refraction is a striking effect in light and electron optics \cite{Veselago,Pendry,Cheianov,Libisch,Betancur}. Junctions acting as metamaterials focus the electron flow towards a spot such as Veselago lens (VL) \cite{Cheianov}. This optics device is claimed to have important uses for controlling particle flow, invisibility cloak \cite{Schurig}, as well as probing tip in a scanning tunneling microscope \cite{Hills}. In light-geometrical optics, conventional diverging and converging lenses are part of multiple optical instruments \cite{Born}. However, the counterpart diverging of a VL has not been proposed yet. This absence can be understood because a more general geometrical optics continues unexplored. The study of anisotropic Dirac materials heterojunctions offers the opportunity of designing that missing lens. 

In this paper it is shown that using ballistic systems with elliptical Dirac cones is possible to obtain super-diverging lens (SDL). The specific condition for a heterojunction formed by one isotropic and other anisotropic Dirac materials is established for redirecting electron flow and creating virtual focus. This perfect lens displays a complete absence of backscattering regardless of the angle of incidence. The omnidirectional conservation of pseudo-spin leads to the first realization of a super-Klein tunneling (SKT) of pseudo-spin 1/2 particles. Such effects emerge because electrons have different refraction laws of pseudo-spin and group velocity when the Dirac cone parameters are changed at the interface. In this way, singular phenomena and novel applications can be achieved. For instance, the use of these superlenses forms an optical device capable of ``seeing" with electrons. The SDLs may be implemented without being necessary a split-gate structure.

\begin{figure*}[t!!]
\centering
\includegraphics[trim = 0mm 0mm 0mm 0mm, scale= 0.4, clip]{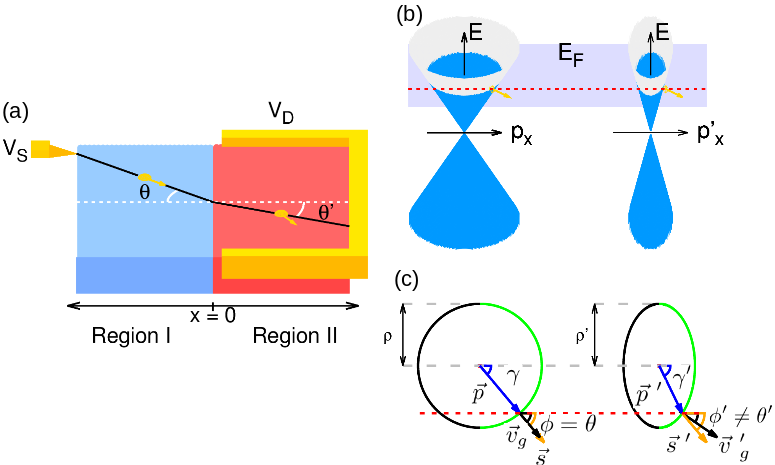} 

\caption{Schematic diagram of massless Dirac fermions refraction in a heterojunction formed by two relativistic Dirac materials. (a) Electrons are emitted by the point source $V_S$. The particle changes its pseudo-spin and group velocity direction at the interface $x = 0$. Then, an extended drain $V_D$ collects the output electron beams. (b) Dirac cone band structure of the heterojunction, where blue solid region shows the occupied states. (c) Kinematical construction illustrates the refraction from the conservation of energy $E$ (circle and ellipse), linear momentum $p_y$ (dashed red line), and probability current density $j_x$ (green semiarcs). The circle (ellipse) is the energy contour at the Fermi level $E_F$ for the region I (II). The refraction index $\rho$ ($\rho'$) is the radius (vertical half-width) of the circle (ellipse). The direction of group velocity (black arrow), pseudo-spin (golden), and linear momentum (blue), are indicated by the angles $\theta$, $\phi$, and $\gamma$, respectively.}
\label{GI}
\end{figure*}

\section{Particle transmission in anisotropic Dirac materials heterojunctions}

Transmission of massless Dirac fermions is considered in a device formed by the junction of relativistic materials, as shown in \fref{GI}(a). The linear interface separates two uniform regions with different anisotropy. This partition is for obtaining one circular (elliptical) Dirac cone in the region I (II), as shown in \fref{GI}(b). Thus, particles impinging on the interface modify the geometry of its dispersion relation. This feature is essentially important for that the particle flow redirects of an unusual way. With this purpose, two-dimensional heterojunctions compose of graphene and uniaxially strained graphene along the zig-zag or armchair direction can be fabricated \cite{Pereira2,Oliva}. Strained artificial systems such as microwave hexagonal lattices \cite{Bellec,Stegmann2}, optical lattices \cite{Shen} and photonic crystals \cite{Louie}, could simulate the particle scattering on elliptical Dirac cone heterojunctions. Three-dimensional case is possible from Weyl and Dirac semimetals \cite{Hills}. The region I can be occupied by one isotropic semimetal, meanwhile the other semimetal might be pressed along the $x$ direction in order to induce anisotropy. 

Ballistic transport is warrantied when the coherence length and mean free path are larger than the device's dimensions \cite{Fuchs,Low2}. Typical experimental values of these quantities are of the order of $\mu$m in graphene and other related materials \cite{Kim,HoLee,Chen}. Longer electron wavelengths are obtained if Fermi level is within low energy regime. Thus, unwanted scattering caused by atomistic details can be avoided. In most of devices, two electrostatic gates $V$ and $V'$ create an abrupt step potential. In this way, angular filter of electron rays beyond the normal incidence is decreased by reduction of evanescent waves \cite{Fuchs,Low2}. All these special conditions have been experimentally achieved in graphene \cite{Kim,HoLee,Chen}. 

In order to describe the scattering of massless Dirac fermions it is applied the Dirac-like Hamiltonian of pseudo-spin 1/2 particles 

\begin{equation}
 H_D = v_F\vec{\sigma}\cdot\vec{p} + V 
\label{HD}
\end{equation}

\noindent for the region I, where $\vec{p}$ is the linear momentum, $v_F$ the Fermi velocity, and $\vec{\sigma} = (\sigma_x,\sigma_y)$ the Pauli matrices. The Hamiltonian \eref{HD} depicts the electron dynamics of isotropic systems such as pristine graphene and related Dirac materials \cite{Castro,Wehling}. The dispersion relation of electrons and holes is written as $E - V = sv_F\sqrt{p^2_x + p^2_y}$, where $s = \textrm{sgn}(E - V)$ is the band index. While the eigenstates are given by the spinor $|\Psi\rangle = (1,s\textrm{e}^{i\phi})\exp(i\vec{p}\cdot\vec{r}/\hbar)/\sqrt{2}$, being $\phi = \arctan(p_y/p_x)$ the pseudo-spin angle. Throughout the text, the unprimed (primed) quantities correspond to the region I (II). In the region II, the Weyl-like Hamiltonian 

\begin{equation}
H'_W = v_F(\lambda'_1\sigma_xp'_x + \lambda'_2\sigma_yp'_y) + V'
\label{HW}
\end{equation}

\noindent is put forward for studying the particle dynamics in anisotropic pseudo-relativistic systems \cite{Armitage,Goerbig}, where  $\lambda'_1 = \cot \alpha'$ ($\lambda'_2 = \cot \beta'$) is related by the extremal elliptical cone angle $\alpha'$ ($\beta'$). These Dirac cone parameters can be obtained from tight-binding approach or DFT calculations \cite{Betancur2}. It is important to note that the Fermi velocity can be always set equal in both sides of the junction by adjusting the extremal angles $\alpha'$ and $\beta'$. Hence, the elliptical Dirac cone is expressed as $E - V' = s'v_F\sqrt{\lambda'^2_1p^2_x + \lambda'^2_2p^2_y}$. The eigenstates $|\Psi'\rangle = (1,s'\textrm{e}^{i\phi'})\exp(i\vec{p}\ '\cdot\vec{r}/\hbar)/\sqrt{2}$ have a different relation of pseudo-spin angle $\phi' = \arctan(\lambda'_2p'_y/\lambda'_1p'_x)$ with the components of the linear momentum. This expression is relevant for establishing the pseudo-spin Snell's law. 

For calculating the transmission probability of electrons crossing the interface, the wavefunction in the region I is $|\Psi_I\rangle = |\Psi\rangle_i + r|\Psi\rangle_r$. The first (second) state on the right side corresponds to the incoming (reflected) electron wavefunction. The coefficient $r$ is the probability amplitude of the reflected electron. The wavefunction in the region II is given by $|\Psi_{II}\rangle = t|\Psi\rangle_t$, where $t$ is the amplitude of the transmitted wave. Using the boundary condition $\Psi_I(\vec{r})|_{x = 0^-} = \xi\Psi_{II}(\vec{r})|_{x = 0^+}$, where $\xi$ is a real quantity which can be obtained by integrating the inhomogeneous Weyl equation \cite{Raoux}, the transmission probability has the form

\begin{equation}
T(\phi,\phi') = \frac{2\cos\phi\cos\phi'}{ss' + \cos(\phi + \phi')},
\label{R}
\end{equation}

\noindent where the conservation of probability current density $j_x$ indicates that $T(\phi,\phi') = ss'|t|^2\cos\phi'/\cos\phi$ instead of $|t|^2$. The complete specification of $T(\phi,\phi')$ in equation \eref{R} must be established by the relation of $\phi$ and $\phi'$. Since the change of Dirac cone geometry in the tunneling makes invalid the use of conventional Snell's law $s|E - V|\sin\phi = s'|E - V'|\sin\phi'$, a novel refraction law is needed in anisotropic Dirac materials. 

\section{Electron refraction laws for the pseudo-spin and group velocity} 

\begin{figure*}[t!!]
\centering
\includegraphics[trim = 0mm 0mm 0mm 0mm, scale= 0.4, clip]{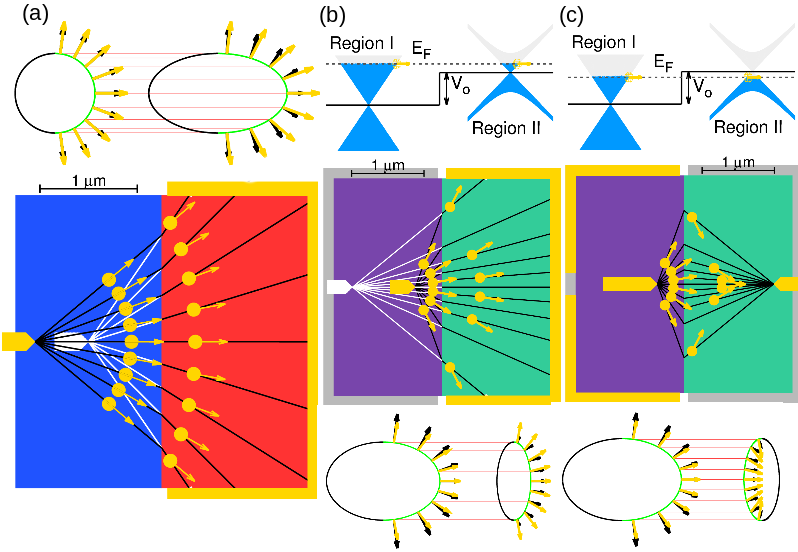} 
\caption{Scattering of electrons in SDLs and VLs. (a) The heterojunction, which is formed by one isotropic Dirac material ($\alpha = \beta = 45^{\textrm{o}}$) and other anisotropic ($\alpha' = 60^{\textrm{o}}$ and $\beta' = 45^{\textrm{o}}$), creates a SDL, where $v_F = 0.83 \times 10^6$ ms$^{-1}$. An electrode source at $x_0 = -1.3\ \mu$m injects electrons with wide angular distribution. The ingoing electron flow, as shown at the time $ t = 1.2 $ ps, is subsequently redirected at the interface to form the virtual focus (white fictitious electrode) at the spot $ x'_0 = -0.75 \ \mu$m. Outgoing electrons at the time $t = 2.3$ ps are plotted. The kinematical construction shows pseudo-spins (golden arrows) which are conserved in the whole incidence range, giving rise to the emergence of a SKT. Meanwhile, the group velocities (black arrows) change the direction and magnitude causing the redirection of divergent flux. (b) [(c)] With a heterojunction presenting two different elliptical Dirac cones, where the values $\alpha = 40^{\textrm{o}}$, $\beta = 30^{\textrm{o}}$, $\alpha' = 50^{\textrm{o}}$, and $\beta' = 70^{\textrm{o}}$ are set, the super divergence (convergence) is reached using two external gates $V = 0$ and $V' = V_0 = 100$ meV. Thus, the Fermi level must be adjusted at the specific value $E_F = 126.6$ ($82.6$) meV given by the diverging (focusing) condition in equation \eref{Esdl} [\eref{Evl}]}.
\label{SDL}
\end{figure*}  

The conservation laws of $E$, $p_y$, and $j_x$, which are schematically represented in \fref{GI}(c), serve for obtaining the specific relationship of $\phi$ and $\phi'$. The crucial point occurs when particles tunnel the elliptical Dirac cone because of that the group velocity, pseudo-spin, and linear momentum have different directions. This important fact gives rise to the appearance of an optical-like phenomena wider than electron isotropic optics, due to that these three quantities satisfy atypical Snell's laws. Pseudo-spin angles $\phi$ and $\phi'$ cannot be interpreted as the angles of incidence and refraction, as usually assumed for isotropic systems \cite{Fuchs,Beenakker}. The genuine angles of incidence $\theta$ and refraction $\theta'$ in anisotropic media are defined by the group velocity. Although electron Snell's law in terms of the pseudo-spin angles (see appendix A)

\begin{equation}
s|E - V|\sin\phi = \frac{s'}{\lambda'_2}|E - V'|\sin\phi',
\label{phi_Snell}
\end{equation}

 \noindent has similar form than isotropic case, it points out singular effects. In equation \eref{phi_Snell} refraction index ratio is written as $n'/n = ss'\rho'/\rho$, where $\rho = |E - V|/v_F$ ($\rho' = |E - V'|/v_F \lambda'_2$) is the radius (vertical half-width) of the circular (elliptical) energy contour at the Fermi level, as seen in \fref{GI}(c). It is interesting to note that setting $V = V'$, the refraction index ratio only depends of the extremal angle $\beta'$. Thus, the pseudo-spin direction is indepedent of the Fermi level. Moreover, when $\rho = \rho'$ and $s = s'$ ($s = -s'$) there is a quite simplification in equation \eref{phi_Snell} obtaining $\phi = \phi'$ ($\phi' = \pi - \phi$). The geometrical criterion $\rho = \rho'$ and $s = -s'$ is the generalization of focusing condition $E = V_0/2$ for a VL in graphene $p$-$n$ junctions. Whereas the other criterion $\rho = \rho'$ and $s = s'$ called as diverging condition indicates the emergence of a novel optics element. If $\phi = \phi'$ in equation \eref{R} is evaluated, then $T(\phi) = 1$. Therefore, electrons always cross the interface regardless of the angle of incidence $\theta$. This effect, which is known as SKT, has been only shown for pseudo-spin one systems \cite{Louie,Bercioux2,Shen,Betancur3}. The present result corresponds to the first prediction of a SKT of pseudo-spin 1/2 particles.

It is worth to identify how massless Dirac fermions are scattered under the conditions $\rho = \rho'$ and $s = \pm s'$. In \fref{GI}(c), propagating modes are linked by the horizontal line which denotes the conservation of $p_y$. Since the current density $\vec{j}$ is directed outward (inward) of the energy contour for $s = 1$  ($s = -1$), electrons can be positively (negatively) refracted for intraband $s = s'$ (interband $s = -s'$) tunneling. Interband tunneling of Dirac materials is the realization in condensed matter of the scattering process of particles turning into antiparticles inside the step potential in high-energy physics. The consideration of anisotropy in Dirac materials heterojunctions modifies the Snell's law, whose expression in terms of $\theta$ and $\theta'$ is given by (see appendix A)

\begin{equation}
s|E - V|\sin\theta = \frac{s'\lambda'_1|E - V'|\sin\theta'}{\lambda'_2\sqrt{\lambda'^2_2\cos^2\theta' + \lambda'^2_1\sin^2\theta'}},
\label{theta_Snell}
\end{equation}

\noindent where the isotropic case is restored when $\lambda'_1 = \lambda'_2 = 1$. For $V = V'$, equations \eref{phi_Snell} and \eref{theta_Snell} are independent of $E$, being unnecessary a split-gate structure. The refraction index ratio has an angular variation which is caused by the anisotropy in the dispersion relation. This important feature is exactly considered in the refraction laws \eref{phi_Snell} and \eref{theta_Snell} for the scattering behavior of electrons in anisotropic Dirac materials heterojunctions. An appropriate characterization of these systems requires to link $\phi'$ and $\theta'$. Thus, the transmission probability \eref{R} as a function of $\theta$ is obtained (see appendix A). 

\section{A novel perfect lens: The super-diverging lens}

\begin{figure}[t!!]
\centering
\begin{tabular}{c}
\includegraphics[trim = 0mm 0mm 0mm 0mm, scale= 0.19, clip]{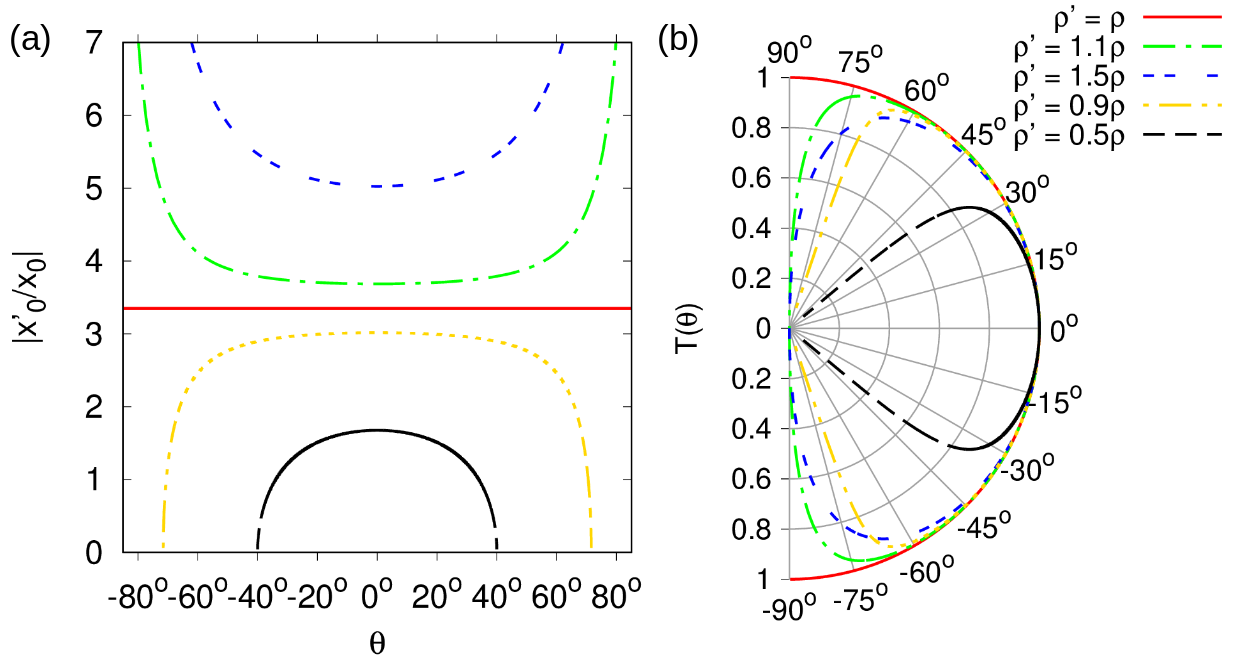}\\
\includegraphics[trim = 0mm 0mm 0mm 0mm, scale= 0.21, clip]{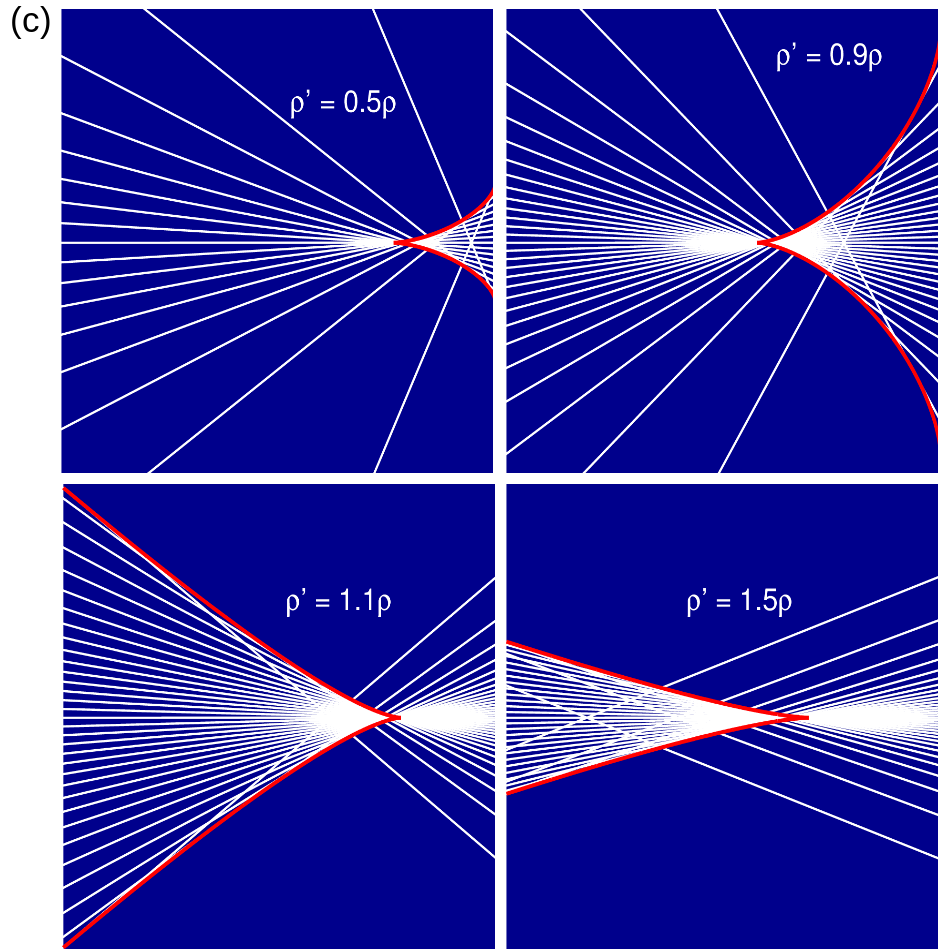}
\end{tabular}
\caption{Deviation of the diverging condition for the doubly anisotropic heterojunction using the same elliptical Dirac cone parameters than in \fref{SDL}(b) and (c). (a) Virtual focus as a function of $\theta$ for $\rho' = 0.5\rho$ (black), $\rho' = 0.9\rho$ (golden), $\rho' = \rho$ (red), $\rho' = 1.1\rho$ (green), and $\rho' = 1.5\rho$ (blue). (b) Probability transmission as a function of $\theta$. If $\rho' \neq \rho$, the omnidirectional perfect tunneling is suppressed for grazing incidence. When $\rho' < \rho$, there is a total internal reflection at the range $|\theta| \geq \theta_c$. (c) Schematic representation of virtual rays with focus angular-dependent. Caustics (red curves) are formed for $\rho' \neq \rho$.}
\label{focus}
\end{figure} 

The application of diverging condition in the Snell's law \eref{theta_Snell} reduces to $\tan\theta' = (x_0/x'_0)\tan\theta$, where $x'_0 = \lambda'_1x_0$ (see appendix B). Heterojunctions with this specific scattering of electrons are shown in \fref{SDL}(a) and (b) where equations \eref{phi_Snell} and \eref{theta_Snell} are used. A point source, which is located at $(x_0,0)$ with $x_0 < 0$, spreads electrons in the whole directions. The group velocity and pseudo-spin have the same direction within the region I. Crossing the interface, the pseudo-spin remains its direction but the group velocity changes. Thus, the outcoming electron flow forms a virtual spot located at $(x'_0,0)$. Then, the SDL and SKT emerge. In light-geometrical optics \cite{Born}, conventional divergent (convergent) lens converts incoming parallel beams, which are emitted by a source at the infinity, to an outgoing diverging (converging) flow. Herein, the SDL always has a virtual spot when the source is located at finite distance of the interface. Likewise, VLs converge the incoming flow for sources with arbitrary location. By these analogies with standard light lenses, the SDL is claimed to be the counterpart of a VL.   

\begin{figure*}[t!!]
\centering
\begin{tabular}{cc}
\includegraphics[trim = 0mm 0mm 0mm 0mm, scale= 0.27, clip]{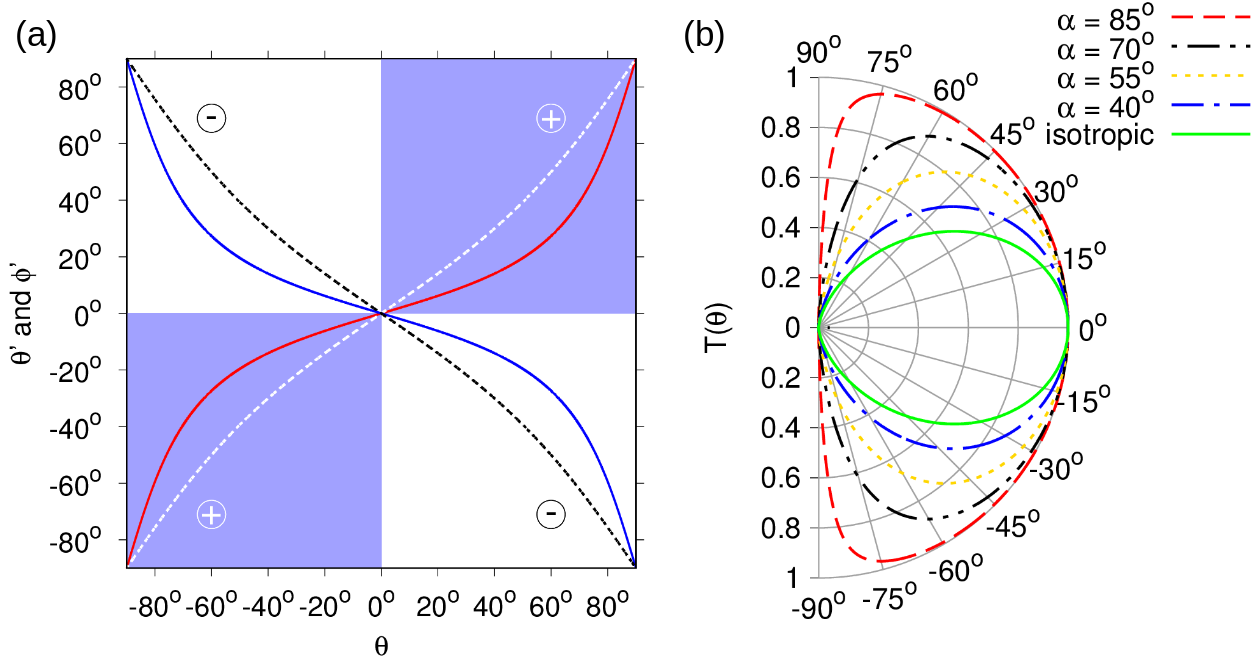}
\end{tabular}
\caption{(a) Angle of refraction $\theta'$ (red and blue line curves) and pseudo-spin direction $\phi'$ (white and black dashed curves) as a function of $\theta$ for the same device in \fref{SDL}(b) and (c). The convention of geometrical optics for angles is used. The blue (white) region indicates positive (negative) refraction for the SDL (VL). (b) Probability transmission as a function of $\theta$ for VLs $pn$-homojunctions using the set of values $\alpha' = \alpha$ and $\beta = \beta' = 30^{\textrm{o}}$. Particle tunneling (dash curves) is enhanced in comparison with the VL of graphene $p$-$n$ junctions (green line curve, $\alpha = \beta = 45^{\textrm{o}}$).}
\label{Snell_transm}
\end{figure*}

So far, the SDL has been proposed for isotropic-anisotropic heterojunctions in absence of a split-gate structure. Thus, the virtual focus occurs regardless of the particle energy and the diverging condition is reduced to $\lambda'_2 = 1$. Notwithstanding, the condition $\lambda'_2 = 1$ could be difficult to obtain in the practice. For doubly anisotropic Dirac materials heterojunctions, the induction of a step potential through the external gates $V = 0$ and $V' = V_0$ does not remove the phenomena. SDL and SKT are reached when the Fermi energy is tuned at the value 

\begin{equation}
E_d = \frac{\lambda_2V_0}{\lambda_2 - \lambda'_2},
\label{Esdl}
\end{equation}

\noindent where $\lambda_1 = \cot\alpha$ and $\lambda_2 = \cot\beta$ are the geometrical parameters of the elliptical Dirac cone in the region I, as shown in \fref{SDL}(b). Such a value is obtained from the diverging condition $\rho = \rho'$ and $s = s'$. It is important to emphasize that SKT and SDL simultaneously emerge for different elliptical Dirac cones in both sides of the heterojunction and tuning the Fermi level at the value given by equation \eref{Esdl}. In order to prove the robustness of the SDL and SKT, the virtual focus for different refraction indexes $\rho$ and $\rho'$

\begin{equation}
x'_0(\theta) = ss'x_0\frac{\lambda'_1\lambda_2\rho'}{\lambda_1\lambda'_2\rho}\sqrt{1 + \left(1 - \frac{\rho^2}{\rho'^2}\right)\frac{\lambda^2_1}{\lambda^2_2}\tan^2\theta}
\label{ang_foc}
\end{equation}

\noindent is calculated (see appendix B). Setting $\rho = \rho'$ in equation \eref{ang_foc}, the virtual focus $x'_0 = x_0\lambda'_1\lambda_2/(\lambda_1\lambda'_2)$ is independent of the angle of incidence $\theta$. This result shows that the super-diverging effect cannot be realized by homojunctions, since the virtual focus matches with the point source. Therefore, the SDLs must be created using anisotropic Dirac materials heterojunctions. One can attempt to find super-diverging flow and omnidirectional perfect transmission examining possible deviated values of the diverging condition. However, the general expression \eref{ang_foc} indicates that the focus is always depending of $\theta$ for $\rho \neq \rho'$, as seen in \fref{focus}(a). Further, the implementation of heterojunctions using isotropic Dirac materials does not prevent the angular dependence of focus. In the case $\rho < \rho'$, the virtual spot is pushed away from the constant one and it has a strong angular dependence for grazing incidence, as seen in \fref{focus}(a). Whereas, the SKT is destroyed in a wide angular range for sizeable deviation in the diverging condition [see \fref{focus}(b)]. Nevertheless, the conservation of pseudo-spin is unaffected for angles near the normal incidence. The lifting of $\rho = \rho'$ causes the formation of virtual caustics with cusp located at $x'_c = x_0\lambda'_1\lambda_2\rho'/(\lambda_1\lambda'_2\rho)$, as shown in \fref{focus}(c). Such an effect also occurs in VLs when focusing condition is lifted off \cite{Cheianov}. The particular shape of caustics in approximated SDLs

\begin{equation}
y_c(x) = \pm\frac{\lambda'_2}{\lambda'_1}\sqrt{\frac{\rho^2(x^{2/3} - x'^{2/3}_c)^3}{\rho'^2 - \rho^2}}
\end{equation}

\noindent is plotted together with the virtual beams in \fref{focus}(c). If $\rho > \rho'$, total internal reflection appears in the range $|\theta| > \theta_c$ where $\theta_c = \arcsin\{[1 + \lambda^2_1(\rho^2/\rho'^2 - 1)\lambda^{-2}_2]^{-1/2}\}$ is the critical angle. The high reflectivity of particles impinging far away the normal incidence favors the robustness of super-divergence, as shown in \fref{focus}(b). This is due to that the rays, whose virtual focuses have an accelerated variation rate on $\theta$, are filtered.

\begin{figure*}[t!!]
\centering
\includegraphics[trim = 0mm 0mm 0mm 0mm, scale= 0.4, clip]{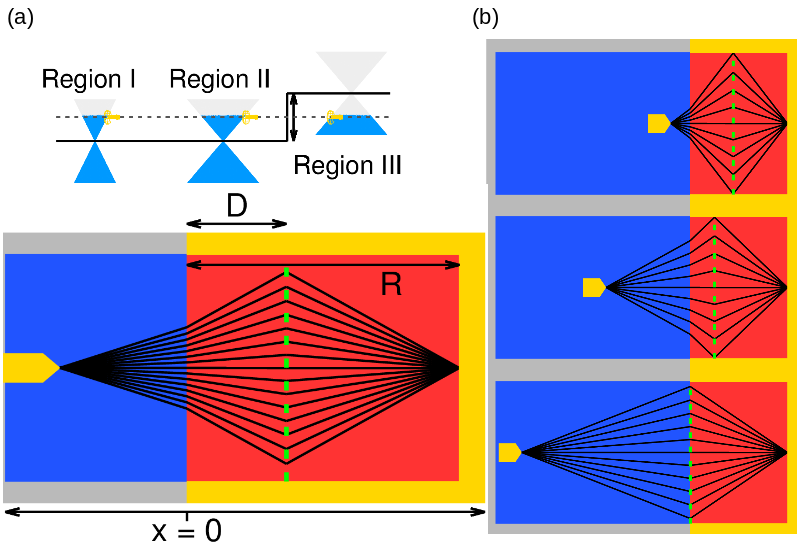}
\caption{Design of an electron eye using two superlenses. (a) The SDL is always located at the interface $x = 0$, while a movable VL appears using the split gate structure at $x = D$, where $V_0 = 100$ meV. The blue (red) region corresponds to the isotropic (anisotropic) Dirac material, where the same set of values in \fref{SDL}(a) are used. Both SDL and VL operate as an eye lens when the regions I and II are negatively doped, while the region III is positively doped at the level $E_F = 50$ meV. (b) Each inset illustrates how the tunable separation between superlenses makes converge the electron flow towards the drain (retina) at $R = 1.5$ $\mu$m. The position of source is changed at $x_0 = -0.3$ (top), $-1.3$ (medium), and $-2.6$ $\mu$m (bottom) and the external gates are moved at $D = 0.66$, $0.38$, and $0$ $\mu$m, respectively.}
\label{e_eye}
\end{figure*}

On the other hand, the same device for obtaining the SDL can also be used as VL [see \fref{SDL}(c)]. Using the focusing condition, the Fermi level must be tuned to the value of 

\begin{equation}
E_f = \frac{\lambda_2V_0}{\lambda_2 + \lambda'_2}
\label{Evl}
\end{equation} 

\noindent and the Snell's law in equation \eref{theta_Snell} is simplified to $\tan\theta' = (x_0/x'_0)\tan\theta$, where $x'_0 = -x_0\lambda'_1\lambda_2/(\lambda_1\lambda'_2)$. Then, the outcoming rays meet at the real focus $(x'_0,0)$ with $x'_0 > 0$. Furthermore, the close connection between both superlenses is better appreciated in \fref{Snell_transm}(a). This special positive refraction suggests a conjugation symmetry with regard to the negative one, which corresponds to the transition from intraband ($s = s'$) to interband ($s = -s'$) tunneling. It is important to mention that the positive refraction given by equation \eref{theta_Snell} substantially differs to the conventional one of gapped graphene $pn$-junctions and other isotropic systems \cite{Dahal}. The present atypical positive refraction allows to create virtual focus independent of $\theta$ and having different location with regard to the point source. These features are essentials for the formation of SDLs and the appearance of SKT.

The transmission probability of VLs, which are designed with anisotropic Dirac materials, exhibits an improved efficiency in comparison with a VL of graphene $p$-$n$ junction [see \fref{Snell_transm}(b)]. In the particular case, where the region I and II have the same anisotropy, the symmetric Veselago lens is recovered. The average transmission $\langle T \rangle = \lambda_2(\lambda_1 + \lambda_2)^{-1}$ enhances in the limit $\lambda_1 << \lambda_2$, doing that $\langle T \rangle$ tends to one, which contrasts with the value of $\langle T \rangle = 0.5$ for circular Dirac cones, as shown in \fref{Snell_transm}(b). With homojunctions of uniaxially strained graphene along the zig-zag direction, VLs can be implemented to obtain high-efficiency transmission. On the other hand, deviations in the focusing condition cause drastical effects in the particle transmission for grazing incidence and lead to the formation of real caustics. This typical aberration in lenses is also appreciated in graphene $p$-$n$ junctions when the focusing condition lifts off \cite{Cheianov}.

\section{The electron optics eye device}

 The use of SDLs and VLs can be taken into account for designing novel electron optics instruments. For instance, a device having both SDL and VL is shown in \fref{e_eye}(a). A movable split gate structure creates the step potential at $x = D$. The action of SDL and VL simulates the eye lens. The tuning of relative separation $D$ between superlenses controls the convergence of rays towards the drain (retina). For a source located at the position $(x_0,0)$, the SDL has a virtual focus at $(x_0\lambda'_1,0)$. In order to focus the refracted electron flow towards the drain at $(R,0)$, where $R$ is the anisotropic Dirac material length [see \fref{e_eye}(b)], the split gate structure needs to be shifted at the position $D = (R + x_0\lambda'_1)/2$. The electron eye loses the focusing ability for sources which are located beyond $x_0 = -R/\lambda'_1$. The increase of relative separation $d$ of the gates produces a smoother step potential causing the angular filter effect of electrons \cite{Fuchs,Low2}. Thus, the operation of pupil is also mimicked. Although three Dirac cones are involved in this system, the transmission probability in equation \eref{R} continues being valid because SKT of electrons is performed within the region I and II. Hence, there is no Fabry-P\'erot interferences.

\section{Conclusions and final remarks}

In summary, the super-diverging lens based on heterojunctions of anisotropic relativistic Dirac materials has been shown. Novel Veselago and super-diverging lenses can be fabricated in two and three-dimensional systems whose electronic band structure presents different Dirac cones parameters in both sides of the junction. The refraction of massless Dirac fermions is governed by a generalized Snell's law in anisotropic media offering advantages in the manipulation of pseudo-spin and guiding of electrons. These novel Snell's laws of pseudo-spin and group velocity allow to calculate the exact direction of outgoing electron beams without considering the approximation of circular Dirac cones in anisotropic media. The special positive and negative refraction of electrons evidence exceptional phenomena, such as super-diverging particles flow, omnidirectional Klein tunneling, and enhanced Veselago lenses, whose control can be of relevant importance in quantum information. The feasibility of designing novel devices using super-diverging lenses may lead to unusual technological applications, where electron eye is a particular example. This new topic called as electron elliptical Dirac optics opens up the possibility of feedback with light-optics in metamaterials. The high efficiency in the particle transmission of anisotropic heterojunctions can considerably improve the operation of well-known optics devices.   

\ack{
Y.B.-O. gratefully acknowledges financial support from CONACYT Proyecto Fronteras 952 Transporte en sistemas peque\~nos, cl\'asicos y cu\'anticos and UNAM-DGAPA-PAPIIT, Project No. IN-103017. The author also thanks to T. Stegmann, F. Leyvraz, T.H. Seligman, G. Cordourier-Maruri, and R. de Coss for helpful discussions, comments, and critical reading of the manuscript.}

\appendix
\setcounter{section}{1}
\section*{Appendix A: Electron Snell's law in anisotropic relativistic Dirac materials}

In order to obtain the different Snell's laws of anisotropic massless Dirac fermions, the parameterization of linear momentum in terms of pseudo-spin $\phi$, angle of incidence $\theta$ or the linear momentum direction $\gamma$ might be found. The expression of $\vec{p}$ as a function of $\phi$  

\begin{eqnarray}
p_x & = & \frac{|E - V|}{v_F\lambda_1}\cos\phi \nonumber\\
p_y & = & \frac{|E - V|}{v_F\lambda_2}\sin\phi,
\label{pphi}
\end{eqnarray}

\noindent is straightforwardly obtained using the wavefunction phase $\tan\phi = \lambda_2p_y/(\lambda_1p_x)$ and the elliptical dispersion relation 

\begin{equation}
|E - V| = v_F\sqrt{\lambda^2_1p^2_x + \lambda^2_2p^2_y}
\label{ell_dr}
\end{equation}

\noindent of anisotropic massless Dirac fermions. Using the conservation of linear momentum $p_y = p'_y$, the electron Snell's law \eref{phi_Snell} in terms of $\phi$ is obtained doing $\lambda_1 = \lambda_2 = 1$ for the circular Dirac cone in the region I. With two different elliptical Dirac cones in both sides of the junction, a more general Snell's law of pseudo-spin is given by

\begin{equation}
s\rho\sin\phi = s'\rho'\sin\phi',
\label{phi_Snell2}
\end{equation}

\noindent being the refraction index ratio $n'/n = s'\rho'/s\rho = ss'|E - V'|\lambda_2/|E - V|\lambda'_2$. An important fact is the control of pseudo-spin direction in heterojunctions without using a split-gate structure. Each relativistic material has a refraction index of $n = \tan\beta$ regardless of the energy. For contrasting the different behavior of pseudo-spin and group velocity in the refraction, it is necessary to calculate the group velocity operator through the Heisenberg equation

\begin{equation}
\hat{\vec{v}} = \frac{i}{\hbar}[\vec{r},H] = v_F(\hat{x}\lambda_1\sigma_x + \hat{y}\lambda_2\sigma_y),
\end{equation}

\noindent where the Weyl-like Hamiltonian \eref{HW} is considered. The components of expected value $\langle \hat{\vec{v}} \rangle$ are given by

\begin{eqnarray}
v_x = v(\theta)\cos\theta  & = & sv_F\lambda_1\cos\phi \nonumber\\
v_y = v(\theta)\sin\theta  & = & sv_F\lambda_2\sin\phi.
\label{v_comp}
\end{eqnarray}
\\
\noindent Thus, inverting the above equation system and substituting in equation \eref{pphi}, the linear momentum as a function of $\theta$

\begin{eqnarray}
p_x & = & \frac{s|E - V|v(\theta)}{v^2_F\lambda^2_1}\cos\theta \nonumber\\
p_y & = & \frac{s|E - V|v(\theta)}{v^2_F\lambda^2_2}\sin\theta.
\label{ptheta}
\end{eqnarray}

\noindent is obtained. Using the equation system \eref{ptheta} and equation \eref{ell_dr}, the group velocity magnitude 

\begin{equation}
v(\theta) = \frac{v_F\lambda_1\lambda_2}{\sqrt{\lambda^2_2\cos^2\theta + \lambda^2_1\sin^2\theta}},
\label{v}
\end{equation}

\noindent is independent of the particle energy but having an elliptical angular variation by the anisotropy in the dispersion relation \eref{ell_dr}. Hence, the conservation of $p_y$ in terms of $\theta$ leads to the Snell's law given by  

\begin{eqnarray}
\frac{s\lambda_1\rho\sin\theta}{\sqrt{\lambda^2_2\cos^2\theta + \lambda^2_1\sin^2\theta}} = \frac{s'\lambda'_1\rho'\sin\theta'}{\sqrt{\lambda'^2_2\cos^2\theta' + \lambda'^2_1\sin^2\theta'}}, & \nonumber \\
& \label{vSnell} 
\end{eqnarray}

\noindent which corresponds to systems with two anisotropic relativistic media. The relation between $\phi$ and $\theta$ can be obtained substituting the group velocity \eref{v} in equations \eref{v_comp} 

\begin{eqnarray}
\cos\phi & = & \frac{s\lambda_2\cos\theta}{\sqrt{\lambda^2_2\cos^2\theta + \lambda^2_1\sin^2\theta}} \nonumber\\
\sin\phi & = & \frac{s\lambda_1\sin\theta}{\sqrt{\lambda^2_2\cos^2\theta + \lambda^2_1\sin^2\theta}}.
\end{eqnarray}

\noindent Therefore, the probability transmission as a function of $\theta$ and $\theta'$ can be expressed as

\begin{eqnarray}
T(\theta,\theta') = & \nonumber\\
\frac{2\lambda_2\lambda'_2\cos\theta\cos\theta'}{\lambda_2\lambda'_2\cos\theta\cos\theta' - \lambda_1\lambda'_1\sin\theta\sin\theta' + f(\theta,\theta')}, & \label{T} \\
 & \nonumber
\end{eqnarray}
 
\noindent where $f(\theta,\theta') = (\lambda^2_2\cos^2\theta + \lambda^2_1\sin^2\theta)^{1/2}(\lambda'^2_2\cos^2\theta' + \lambda'^2_1\sin^2\theta')^{1/2}$. Equation \eref{T} is the analog of Fresnel coefficient in the electromagnetic theory \cite{Born}. On the other hand, a third Snell's law in terms of the linear momentum angle $\gamma$ holds. Using the parameterization $p_x = p(\gamma)\cos\gamma$, $p_y = p(\gamma)\sin\gamma$, and evaluating in equation \eref{ell_dr}, it is possible to show that

\begin{eqnarray}
\frac{\lambda_2\rho\sin\gamma}{\sqrt{\lambda^2_1\cos^2\gamma + \lambda^2_2\sin^2\gamma}} = \frac{\lambda'_2\rho'\sin\gamma'}{\sqrt{\lambda'^2_1\cos^2\gamma' + \lambda'^2_2\sin^2\gamma'}}, & \nonumber\\
 & \label{pSnell}
\end{eqnarray}

\noindent where the conservation of $p_y$ is again used. The three refraction laws \eref{phi_Snell2}, \eref{vSnell}, and \eref{pSnell} are reduced to the standard form for the case of circular Dirac cones. The Snell's laws for the linear momentum and group velocity are very similar, they can be related using the substitution $\theta \rightarrow \gamma$, interchanging the $\lambda$ parameters $\lambda_1 \rightarrow \lambda_2$, $\lambda_2 \rightarrow \lambda_1$ (also primed quantities), and omitting the band index.  

\appendix
\setcounter{section}{2}
\section*{Appendix B: Refraction law of superlenses and angular dependence of focus}

A heterojunction formed by two anisotropic relativistic Dirac materials works as a superlens if the electron optics conditions are fulfilled. Any incident particle emitted by a point source, which is located at $(x_0,0)$, follows the ray equation 

\begin{equation}
y = (x - x_0)\tan\theta,
\end{equation}

\noindent being valid in the range $x_0 \leq x \leq 0$. Using the primed version of equation \eref{ptheta}, the elliptical dispersion relation \eref{ell_dr}, and conservation $p_y = p'_y$, the refracted particles have the group velocity direction

\begin{equation}
\tan\theta' = \frac{v'_y}{v'_x} = \frac{s'\lambda'_2p_y}{\lambda'_1\sqrt{\rho'^2 - p^2_y}}.
\end{equation}

\noindent A similar expression $\tan\theta = s\lambda_2p_y/(\lambda_1\sqrt{\rho^2 - p^2_y})$ is also satisfied for incident particles. Since the common condition between Veselago and super-diverging lenses is $\rho = \rho'$, the expression of Snell's law for both electron optics devices can be written as

\begin{equation}
\tan\theta' = ss'\frac{\lambda_1\lambda'_2}{\lambda'_1\lambda_2}\tan\theta.
\end{equation}

\noindent Then, the ray equation in the region II ($x > 0$) can be reduced to

\begin{eqnarray}
y & = & x\tan\theta' - x_0\tan\theta \nonumber \\
y & = & \left(ss'\frac{\lambda_1\lambda'_2}{\lambda'_1\lambda_2}x - x_0\right)\tan\theta,
\end{eqnarray}

\noindent showing that the outcoming electron flow meets in a real ($s = -s'$) or virtual ($s = s'$) focus given by  

\begin{equation}
x'_0 = ss'x_0\frac{\lambda'_1\lambda_2}{\lambda_1\lambda'_2}.
\label{xpo}
\end{equation}

\noindent This result is very important for obtaining a superlens because the focus does not depend of $\theta$. The expression \eref{xpo} can also be derived from the general focus equation 

\begin{eqnarray}
x'_0(\theta) & = & x_0\frac{\tan\theta}{\tan\theta'} = ss'x_0\frac{\lambda'_1\lambda_2}{\lambda_1\lambda'_2}\sqrt{\frac{\rho'^2 - p^2_y}{\rho^2 - p^2_y}} \nonumber \\
 & = & ss'x_0\frac{\lambda'_1\lambda_2\rho'}{\lambda_1\lambda'_2\rho}\sqrt{1 + \left(1 - \frac{\rho^2}{\rho'^2}\right)\frac{\lambda^2_1}{\lambda^2_2}\tan^2\theta} 
\end{eqnarray}

\noindent doing $\rho = \rho'$. The focus has an angular dependence when $\rho \neq \rho'$, as shown in \fref{focus}, and it presents drastic angular variation for grazing incidence.

\end{document}